\documentstyle[aps,twocolumn,floats]{revtex}

\input epsf

\makeatletter
\global\@specialpagefalse
\def\@oddhead{\small \sl Giant Thermomechanical Effect in Normal Liquid $^3$He\hfill
 submitted to Canadian Journal of Physics}
\let\@evenhead\@oddhead

\def\@oddfoot{\reset@font\rm\hfill \thepage\hfill

} \let\@evenfoot\@oddfoot
\makeatother

\begin{document}

%\draft

\wideabs{
\title{Giant Thermomechanical Effect in Normal Liquid $^3$He}
\author{D.L. Sawkey, D. Deptuck, D. Greenwood, and J.P. 
Harrison\cite{JPH} }
\address{Physics Department, Queen's University, Kingston, Ontario K7L 3N6,
Canada}
\date{September 30, 1997}
\maketitle
\begin{abstract}
Measurements are presented of the thermomechanical coefficient of normal
liquid $^3$He confined in a porous plug pre-plated with four monolayers
of $^4$He. These nonmagnetic monolayers displace the magnetic solid-like
$^3$He monolayers that are adjacent to the pore surfaces when the plug
is filled with pure $^3$He. In the low temperature limit $(T \leq $ 10
mK) the coefficient can be described by $\Delta P/\Delta T \sim s/6v$
where $\Delta P$ is the pressure difference across the plug generated by
the temperature difference $\Delta T$ and $s$ and $v$ are the molar
entropy and molar volume. This low temperature limit corresponds to the
condition $d \ll \ell_q$ where $d$ is the pore diameter and $\ell_q$ is
the bulk liquid $^3$He quasiparticle mean free path; that is, the
quasiparticles are predominantly boundary scattered in the pores. The
measured coefficient is half that calculated by Edwards, Culman and He.
When compared with this new experimental result for the $^4$He-plated
porous plug, the earlier result for pure liquid $^3$He is strikingly
larger (by up to 30$\times$ at 2~mK). This enhancement is reminiscent of
the giant thermopowers measured in Kondo and other dilute magnetic
alloys. It is speculated that the enhanced thermomechanical coefficient
for pure liquid $^3$He is due to magnetic scattering of the $^3$He
quasiparticles by the two magnetic solid-like $^3$He monolayers adjacent
to the pore surfaces.

\vspace{.1in}
\end{abstract}
}

%\narrowtext

\section{Introduction}

An earlier paper presented results for a new thermomechanical effect, in
normal liquid $^3$He confined within the nanoscale pores of a porous
plug \cite{1}. This was seen as the liquid $^3$He Fermi liquid analogue
of thermoelectricity, the classical gas thermomolecular effect, and the
phonon thermomechanical effect \cite{Andreev} in a general liquid. There
is also the analogy with the superfluid $^4$He thermomechanical effect,
except that this is a reversible effect\cite{2} (no entropy production
or flow) whereas the others need to be described by irreversible
thermodynamics \cite{3,4}. The thermomechanical effect in normal liquid
$^3$He required boundary scattering of the $^3$He quasiparticles just as
the thermomolecular effect in a classical gas and the phonon
thermomechanical effect need an orifice or tube with diameter less than
the molecular and phonon mean free paths, respectively.

The measurement itself was of $\Delta P/\Delta T$ where $\Delta T$ was
the temperature difference across the plug and $\Delta P$ was the
resulting pressure difference.\footnote[2]{Historically, $\Delta
P/\Delta T$ was called the thermomolecular effect for the classical gas
and the thermomechanical effect for superfluid $^4$He. We have chosen to
use the term thermomechanical effect because we are studying liquid
helium, although in the normal state.} The temperature difference was
established by a heat current through the $^3$He confined within the
pores of the plug and measured with a cerous magnesium nitrate
thermometer, calibrated against a $^3$He melting-curve thermometer
\cite{5}. The pressure was measured absolutely in terms of the head of
liquid $^3$He in a liquid $^3$He manometer. The thermomechanical
coefficient $\Delta P/\Delta T$ was substantial, $\sim$ 10 kPa/K or
$\sim$ 10~mm $^3$He/mK over the temperature range 2--25~mK.

A separate experiment showed that the thermomechanical coefficient was
zero for an open geometry, a tube with diameter $D \gg \ell_q$, where
$\ell_q$ is the $^3$He quasiparticle mean free path for scattering by
other quasiparticles. This is also the case for the classical gas
thermomolecular effect in the hydrodynamic limit $D \gg \lambda$, where
$\lambda$ is the molecular mean free path \cite{3}. The measured liquid
$^3$He coefficient with the confined $^3$He was therefore some weighted
average of the boundary scattering coefficient and the zero coefficient
due to scattering by other quasiparticles. A Nordheim-Gorter\cite{6}
procedure, familiar in the field of thermoelectricity, was used to
extract the coefficient due to boundary scattering alone. This
correction was important for $T \geq$~10~mK where $\ell_q \leq d$ and
$d$ is the pore diameter. 

 The introduction continues with two
sections which describe a comparison of the measured coefficient with a
theoretical calculation and a review of some surface effects that could
account for the resulting disagreement.

\subsection{Theory}

The first theoretical approaches to thermoelectricity in a metal and the
thermomolecular effect in a gas were based upon equilibrium
thermodynamics (see, for example, the monograph by Denbigh\cite{4}).
Consider the case of two chambers containing a classical gas connected
by a small orifice; chambers 1 and 2 are at temperatures $T$ and $T +
\Delta T$ and pressures $P$ and $P+\Delta P$. In dynamic equilibrium the
effusion rates from 1 to 2 and from 2 to 1 are the same. Elementary
thermodynamics and kinetic theory then lead to $\Delta P/\Delta T =
-q^\ast /vT$ and $q^\ast = -cT/3$, respectively, where $q^\ast {,} ~ v$
and $c$ are the molar heat of transport, molar volume, and constant
volume molar specific heat respectively. The heat of transport is a flow
of heat from hot to cold, while in dynamic equilibrium, and hence is a
continuing source of entropy. The process is therefore irreversible and
irreversible thermodynamics must be used.

The linear transport laws within irreversible thermodynamics can be
written in terms of generalized currents and forces as $J_i = \sum_k
L_{ik}X_k$ where $L_{ik}$ is a transport coefficient and $J_i {,} ~ X_i$
are chosen so that $\sum_i J_iX_i = dS/dt$, the rate of entropy
production \cite{3}. Onsager\cite{7} used the principle of microscopic
reversibility to show that $L_{ij} = L_{ji}$. In our case, the driving
forces for the liquid $^3$He quasiparticles arise from the differences
of chemical potential, $\mu$, and of temperature across the porous plug.
With quasiparticle molar flow rate $J_N = dN/dt$ and energy flow rate
$J_U = dU/dt$, Onsager's equations become
\begin{equation}
J_N = -L_{11}\Delta \left( {\mu\over T} \right) + L_{12}\Delta \left(
 {1\over T} \right)
\end{equation}
\begin{equation}
J_U = -L_{21}\Delta \left( {\mu\over T} \right) + L_{22}\Delta \left(
 {1\over T} \right)
\end{equation}
or, using standard thermodynamics identities,
\begin{equation}
J_N = - {L_{11}v\over T}\Delta P + {L_{11}h - L_{12}\over T^2}\Delta T
\end{equation}
\begin{equation}
J_U = -{L_{21}v\over T}\Delta P + {L_{21}h - L_{22}\over T^2}\Delta T
\end{equation}
where $h$ is the molar enthalpy.  
The thermomechanical coefficient results from the condition $J_N = 0$, 
so that
\begin{equation}
{\Delta P\over\Delta T} = {h - (L_{12}/L_{11})\over vT}
\label{eq:5}
\end{equation}
As an aside, the condition $\Delta T = 0$ gives
\begin{equation}
J_U = {L_{21}\over L_{11}}J_N \equiv u^\ast J_N
\end{equation}
where $u^\ast$ is the molar energy of the transport.  Therefore,
equation (\ref{eq:5}) becomes
\begin{equation}
{\Delta P\over\Delta T} = {h - u^\ast\over vT} \equiv -{q^\ast\over vT}
\end{equation}
where $q^\ast = u^\ast - h$ is the heat of transport.  Thus, the 
reversible thermodynamics result is retrieved.  To go further, one
needs a microscopic model for the system of interest, for example
kinetic theory for the classical gas.

Following the formalism used by Smith in his review \cite{8} for
transport in a Fermi liquid, Edwards, Culman and He\cite{9} developed
the particle current and energy current transport equations for
simultaneous pressure and temperature gradients. Both cylindrical
geometry and a mesoscopic structure described in terms of the
Landauer-Buttiker formalism were considered. In both cases Edwards et
al.\ found
\begin{equation}
{\Delta P\over \Delta T} = {s\over 3v} = {C_v\over 3}
\end{equation}
where $s$ and $v$ are the molar entropy and molar volume, and $C_v$ is
the constant volume heat capacity per unit volume. The second equality
results from the linear temperature dependence of the heat capacity. The
positive sign signifies a flow of $^3$He from cold to hot in
establishing the pressure difference. Note that $C_v/3$ is identical to
the thermomolecular coefficient in a classical gas and $s/3v$ is 1/3 the
thermomechanical effect in a superfluid. 

Compared to the theoretical result, the measured coefficient was larger
by 15$\times$ at 2 mK and 3$\times$ at 20 mK, although the sign was
correct. Disagreements of this size, and larger, are well known in
thermoelectricity \cite{10}. In general they are associated with energy
dependent scattering by Kondo and other magnetic impurities \cite{11}.

\subsection{Boundary Scattering of $^3$He Quasiparticles}

For liquid $^3$He, the obvious source of magnetic scattering is the
magnetic solid-like layer of $^3$He atoms adjacent to the $^3$He-copper
oxide interface within the porous plug; another
possibility\footnote[3]{Julian Brown, private communication.} is the
magnetism of the copper oxide itself\cite{13}. A Curie-law component in
the magnetism of confined liquid $^3$He was first found for $^3$He in
vycor glass \cite{14}. Curie-Weiss behaviour was demonstrated a few
years later for $^3$He confined between mylar sheets \cite{15}. The
magnetism was clearly identified with $^3$He adjacent to the interface
by pre-plating the mylar with two monolayers of $^4$He atoms and finding
that the Curie-Weiss term was absent. The $^4$He atoms are
preferentially attracted to the interface\cite{16,17}; although the van
der Waals force of attraction to the interface is identical for $^3$He
and $^4$He, the larger mass of the $^4$He atom leads to a lower ground
state energy and consequently a larger binding energy. The first two
monolayers adjacent to the interface are bound with high density and are
solid-like whereas the third and higher monolayers are part of the bulk
liquid. Therefore the pre-plating of two or more monolayers of $^4$He
leaves all of the $^3$He as liquid with the small Pauli paramagnetism of
a Fermi liquid.

By now there is a fairly clear picture of the $^3$He interface
magnetism\cite{18,19,20} and, to a lesser extent, of the scattering of
$^3$He quasiparticles at the boundary either with pure $^3$He or with
$^4$He pre-plating\cite{21,22,23,24}. For pure $^3$He on very flat
carbon (Grafoil) substrates the first monolayer has solid-like density
and is a Curie-law paramagnet. The binding to the underlying substrate
is sufficiently strong that there is no significant exchange within the
monolayer. The second monolayer is less dense, but still solid-like. It
is a Curie-Weiss paramagnet with Weiss constant $\theta \sim -0.5$ mK
(antiferromagnetic). The magnetism of the second monolayer is
independent of whether the first monolayer is $^3$He or $^4$He,
suggesting that the second layer exchange is intra-layer. As this second
monolayer is overlaid with more $^3$He, $\theta$ rapidly switches from
$\sim -0.5$~mK to $\sim + 0.5$~mK (ferromagnetic), although whether this
is caused by interlayer exchange or density enhancement of the second
monolayer is not known. Bozler et al.\cite{25} have shown evidence of
ferromagnetic order of the solid-like $^3$He below 1~mK by measuring the
magnetization in a very small applied field. On other substrates such as
mylar\cite{15,22}, fluorocarbon beads\cite{26}, sintered silver
powder\cite{27}, and aerogel\cite{28} (a silica structure with up to
99\% porosity) there is also Curie-Weiss paramagnetism with $\theta \sim
+ 0.5$ mK, although specific heat measurements\cite{29,30} suggest that
a wide range of $\theta$ values is required to explain the results. In
all cases where the substrates have been pre-plated with two or more
monolayers of $^4$He, the surface magnetism has been
eliminated\cite{15,22,27,28}.

Evidence for the nature of the scattering of $^3$He quasiparticles at
the interface, whether pure $^3$He with its surface magnetism or $^4$He
pre-plated, has come from a variety of experiments: the drag force on
moving $^3$He, the boundary value of the superfluid order parameter,
$^3$He spin relaxation and energy exchange across the interface (Kapitza
conductance). Viscous slip at the boundary is an important correction to
measurements of viscosity\cite{31}. It can be described by a specularity
constant $\nu$, the fraction of quasiparticles that are specularly
scattered at the boundary; the rest are diffusely scattered\cite{8}.
This is a concept that goes back to Maxwell\cite{32}. Specularity
enhances flow through a tube by a factor $(1 + \nu)/(1 - \nu)$. Ritchie
et al.\cite{21} measured the temperature dependent response of a Stycast
epoxy torsional oscillator filled with $^3$He and found a decrease in
the real and imaginary components of the transverse surface impedance
when $\sim$ 2 or more $^4$He monolayers were pre-plated onto the epoxy,
clear evidence for increased specularity. However, they were unable to
fit their results with a theoretical model. Freeman and Richardson\cite{22},
working with a torsional oscillator with a mylar surface, found similar
behaviour and specularities consistent with the range 0.75--0.9 with
$^4$He monolayers present. Tholen and Parpia\cite{33} similarly measured
a jump in $\nu$ from 0.4 to 0.9 when two $^4$He monolayers were added to
a silicon torsional oscillator.

Freeman and Richardson\cite{22} and Steel et al.\cite{24} studied superfluidity
in thin films of $^3$He, on mylar and copper substrates respectively.
Both groups found suppression of the transition temperature for pure
$^3$He; the suppression was consistent with zero order parameter at the
interface, as expected for diffuse scattering of the $^3$He
quasiparticles. Surprisingly at the time, the two groups found no
suppression of the transition temperature when the mylar or copper
substrates were pre-plated with $\sim$ 2 monolayers of $^4$He,
suggesting that the $^4$He-plated interface was acting as a mirror
surface for the $^3$He quasiparticles. Kim et al.\cite{23} found similar
behaviour from fourth sound measurements on $^3$He confined in a packed
powder: suppression of the superfluid density with pure $^3$He and a
relative increase when the pores were pre-plated with $^4$He. Ritchie et
al.\ and Freeman and Richardson have emphasized the dilemma of these results:
The surfaces are not atomically flat and will not be made so by the
addition of just two monolayers of $^4$He. Therefore, geometrically
diffuse scattering is to be expected for both $^3$He and $^4$He at the
interface. 

Other experiments have addressed the question of magnetic spin flip and
energy exchange at the boundary. In general the low temperature
relaxation of magnetization in liquid $^3$He is determined by relaxation
at the boundary. This was demonstrated by Kelly and Richardson\cite{34},
Hammel and Richardson\cite{26}, and Godfrin et al.\cite{35} among
others. In particular, by writing $T_1 = d/\epsilon v_{\rm F}$, where
$T_1$ is the spin relaxation time, $d$ is the characteristic size of the
pores and $\epsilon$ is the spin-flip probability at the
boundary\cite{36}, Godfrin et al.\ found $\epsilon \sim 10^{-6}$ for
$^3$He confined in the pores of platinum powder and $\epsilon \sim
10^{-8}$ in alumina powder and Grafoil. These values of $\epsilon$ were
reduced by a factor $\sim$ 100 when the substrates were pre-plated with
2.7 monolayers of $^4$He; a similar observation had been made by Kelly
and Richardson. In contrast with these very low probabilities for spin
flip, NMR measurements of $^3$He in confined geometries show just one
absorption line\cite{Schuhl,27,28}. The frequency is a weighted average
of the solid and liquid $^3$He frequencies and reflects the rapid
exchange of $^3$He atoms between the liquid $^3$He and solid surface
$^3$He atoms.

The probability for energy exchange by a $^3$He quasiparticle can be
deduced from the Kapitza thermal boundary resistance\cite{37,38}. In his
review Harrison concluded that a lower limit had been reached at $RT
\sim$ 300 K$^2$/W for 1 cm$^3$ of sintered metal powder heat exchanger.
This can be re-expressed as $\dot{Q}/TV\Delta T \sim 3 \times 10^3$ W
m$^{-3}$ K$^{-2}$ where $\dot{Q}/\Delta T$ is the boundary conductance
and $V$ is the volume of the heat exchanger. A theoretical
model\cite{39} for the heat exchange gave the result $\dot{Q}/TV\Delta T
\sim 4 \times 10^{-15}/d^3$ W m$^{- 3}$ K$^{-2}$ which was shown to give
a reasonable representation of the experimental results\cite{40}. This
result translates into a probability $\epsilon^\prime$ for thermal
energy transfer by a quasiparticle at the boundary of $\epsilon^\prime
\sim 10^{-10}$ for 1~$\mu$m pores and $\epsilon^\prime \sim 10^{-8}$ for
0.1 $\mu$m pores.

The conclusions to be drawn from the experiments are that for pure
$^3$He there is a Curie-Weiss surface magnetism which is eliminated by
pre-plating the substrate with two or more $^4$He monolayers. The pure
$^3$He surface scatters the $^3$He quasiparticles largely diffusely but
with a very small probability of spin-flip or energy exchange. Without
the surface magnetism the $^3$He quasiparticles are scattered
specularly, or largely specularly, and the probability of spin-flip or
energy exchange is reduced even further. Following Freeman and Richardson
\cite{22}, since two monolayers cannot flatten a rough surface, the
surfaces must be atomically flat, at least over a length scale
comparable to the de Broglie wavelength of the $^3$He quasiparticles
($2\pi/k_{\rm F} \sim 1$~nm). Therefore the diffuse scattering at the
magnetic surface must be magnetic in origin, but not, in general,
accompanied by spin-flip or energy exchange. Sprague et al.\cite{28}
have considered a magnetic scattering model based upon the scattering of
quasiparticles by a field induced magnetic polarization of the surface
$^3$He atoms, but a general calculation is still required.

In view of the above discussion, it is clear that the important test for
attributing enhancement of the thermomechanical effect to diffusive
magnetic scattering at the interface is to pre-plate the porous plug
with two or more monolayers of $^4$He atoms. A reason for expecting that
this would have an effect goes back to the original observation\cite{24}
that led us to the thermomechanical effect in liquid $^3$He: During
adiabatic demagnetizations of the PrNi$_5$ nuclear cooling refrigerator
there was evidence for movement of $^3$He in runs with pure $^3$He but
not in 24 of the 27 runs with $^4$He pre-plating.

This paper presents the results of pre-plating the porous plug with four
monolayers of $^4$He. Section 2 presents a summary of the experiment.
Section 3 is a presentation and discussion of the results, and
conclusions are drawn in Section 4. A brief description of the results
has been presented at the 1997 Symposium on Quantum Fluids and
Solids\cite{41}.

\section{The Experiment}

\begin{figure}
\epsfxsize=3.2in
\centering{\epsfbox{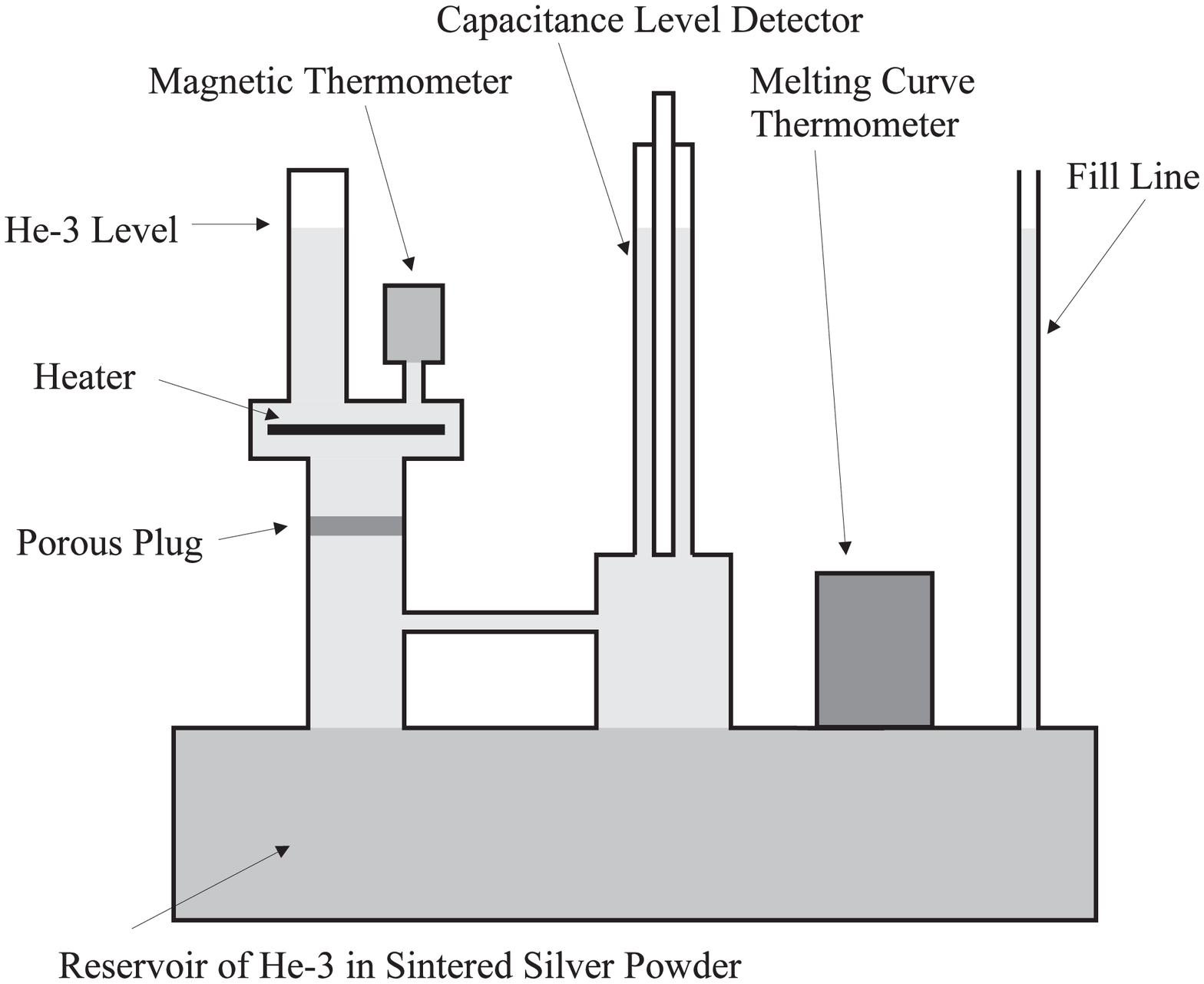}}
\vspace{.05in}
\caption{A schematic illustration of the apparatus.  The left-hand tower
and the capacitance level detector, together with the cross-over tube,
form a liquid $^3$He manometer.  The $^3$He chamber is connected below
to the adiabatic demagnetization refrigerator and above, via a
superconductor heat switch, to a dilution refrigerator.}
\label{fig1}
\end{figure}

The apparatus and experiment have been described elsewhere\cite{1}. For
completeness, a schematic diagram and brief description are included
here. The U-tube geometry of the $^3$He manometer is shown in Fig.\
\ref{fig1}. The bottom of the U-tube was open to the $\sim$ 10 cm$^3$
reservoir of liquid $^3$He in the heat exchanger attached to the
adiabatic demagnetization refrigerator. The left-hand arm or tower
contained a thin film heater, thermally isolated from the apparatus by
thin superconducting leads, and a cerous magnesium nitrate magnetic
thermometer. It also contained the porous plug which separated the
$^3$He into two regions, ideally at temperatures $T + \Delta T$ above
the plug (with added heat) and $T$ below the plug. In practice, there
were temperature gradients within the $^3$He and, as described below,
corrections had to be made. The right-hand tower was a coaxial capacitor
level detector, partially filled with liquid $^3$He also at temperature
$T$. Once the apparatus had been filled with liquid $^3$He and
equilibrium established, pressure differences across the plug were
indicated by level changes in the coaxial capacitor. Short term
(minutes) and long term (hours) sensitivities of 1 and 5~$\mu$m were
realizable. The level detector was calibrated by measuring the empty and
full capacitance values and measuring separately the length of the
capacitor.

The porous plug was packed 70 nm oxidized copper powder, 6.3~mm diameter
by 1.3~mm height. The packing fraction was 0.35 $\pm$ 0.03 by volume.
Oxidized copper was used so that the thermal conductance of the plug
would be negligible compared to that of the liquid $^3$He in the pores.
The mechanical time constant for $^3$He flow through the porous plug,
and hence of the manometer, was long and the consequent 1--2 days
required for each data point was an obvious disadvantage compared to a
diaphragm pressure transducer\cite{42}; however, the level detector does
have the advantage of giving an absolute measure of the pressure
difference.

A separate tower above the main heat exchanger chamber contained a
$^3$He melting curve thermometer; this was used to monitor the
temperature of the $^3$He below the porous plug and, with no added heat,
to calibrate the magnetic thermometer.

The experiment proceeded as follows: The $^4$He for the pre-plating,
equivalent to four monolayers, was admitted to the $^3$He space at
liquid nitrogen temperature and given time at that temperature and also
at 4~K and 1~K to distribute itself uniformly over the $\sim$ 50 m$^2$
surface. The melting curve thermometer was filled at 1~K and the $^3$He
space partially filled (see Fig.~\ref{fig1}) at $\sim$ 0.5~K where the
$^3$He viscosity is a minimum. The lower stage was cooled to below 1~mK
and gradually warmed to calibrate the melting curve thermometer against
the superfluid $^3$He A-transition and to calibrate the magnetic
thermometer over the range 1.5--25 mK. A second cycle was then used to
cool the lower stage to 1.5 mK where the measurements were started.

Ideally, the $^3$He level needed 20--30 hours to settle; however, the
routine filling of the main helium dewar every 36 hours was disruptive
to the level and therefore a complete measurement was made within this
interval. The measurement consisted of three stages: The level was
allowed to settle for $\sim$ 8 hours, heat was added above the porous
plug for 8--10 hours, and finally the heat was switched off. Throughout
the interval, the level was monitored, indicating the initial approach
to equilibrium, the approach to the level difference generated by the
added heat, and the final approach to equilibrium. In all cases the
final levels had to be established by extrapolation. Each level change
was small and consequently it was difficult to determine both the final
level and time constant with any accuracy. Therefore during one of the
intervals between helium fills a large level change was induced and
allowed to decay; this fixed the time constant which is temperature
independent up to 20 mK\@. Each final level was then determined by
making several fits of small sections of each level versus time curve to
an exponential decay differential equation with the time constant fixed,
and averaging. Thermal equilibrium was not a problem; the magnetic
thermometer time constant was $\sim$ 10 minutes.

The result of the experiment itself was a tabulation of $\Delta T$ and
$\Delta H$ for the particular heat input $\dot{Q}$ used at temperature
$T$, for a set of temperatures from 1.5~mK to 20~mK. The $^3$He level
difference, $\Delta H$, was converted to $\Delta P$ by correcting for
the small level change above the plug (see Fig.\ \ref{fig1}) where the
cross-section was 20$\times$ larger than that within the capacitor and
then multiplying by the usual $\rho$g.

\section{Results and Analysis}

\begin{figure}
\epsfxsize=3.2in
\centering{\epsfbox{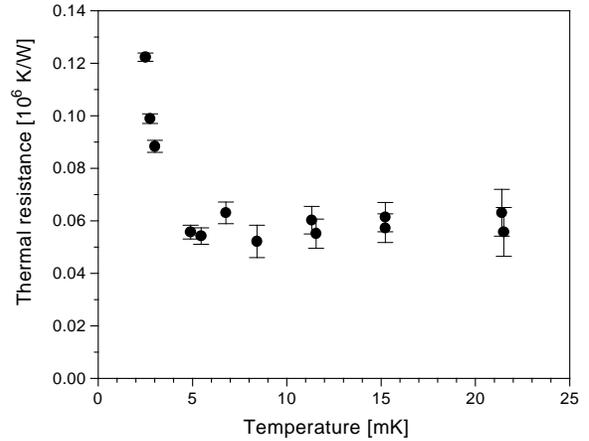}}
\caption{The measured thermal resistance of the liquid $^3$He between
the thin film heater and the $^3$He reservoir.}
\label{fig2}
\end{figure}

The thermal resistance, $R = \Delta T/\dot{Q}$, is shown as a function
of temperature in Fig.~\ref{fig2}. It has a temperature dependence that
reflects the dominance of boundary scattering of the $^3$He
quasiparticles at low temperature and of scattering by other
quasiparticles at high temperature. 
An analysis of these measurements
must take into account the thermal resistance of the $^3$He in the
porous plug, where scattering is by both boundaries and other
quasiparticles and the thermal resistance of the bulk $^3$He above and
below the plug, where boundary scattering can be neglected. Within Fermi
liquid theory, the thermal conductivity of liquid $^3$He is given
by\cite{8}
\begin{equation}
\kappa = {1\over 3}Cv_{\rm F} \ell
\label{eq:9}
\end{equation}
where $C \propto T$ is the heat capacity per unit volume, $v_{\rm F}$ is
the Fermi velocity, and $\ell$ is the $^3$He quasiparticle mean free
path. In the bulk $\ell = \ell_q \propto T^{-2}$, the mean free path due
to scattering by other quasiparticles. In the plug, in the usual way
with two scattering processes, $\ell = \ell_qd/(\ell_q + d)$ where $d$
is the pore diameter. Therefore the total thermal resistance is
\begin{equation}
R = R_{\rm plug} + R_{\rm bulk}
\end{equation}
\begin{equation}
~~=\left( {L\over\kappa A} \right)_{\rm plug} + \left( {L\over 
\kappa A} \right)_{\rm bulk}
\label{eq:11}
\end{equation}
\begin{equation}
~~=\alpha^\prime T  \left(1 + {\ell_q\over d} \right) + 
\alpha^{\prime\prime} T
\label{eq:12}
\end{equation}
\begin{equation}
~~=\alpha T + \beta /T
\end{equation}
where $\alpha^\prime {,}~\alpha^{\prime\prime} {,}~\alpha$ 
and $\beta$ are
constants.  That is, we expect $RT = \alpha T^2 + \beta .$

\begin{figure}
\epsfxsize=3.2in
\centering{\epsfbox{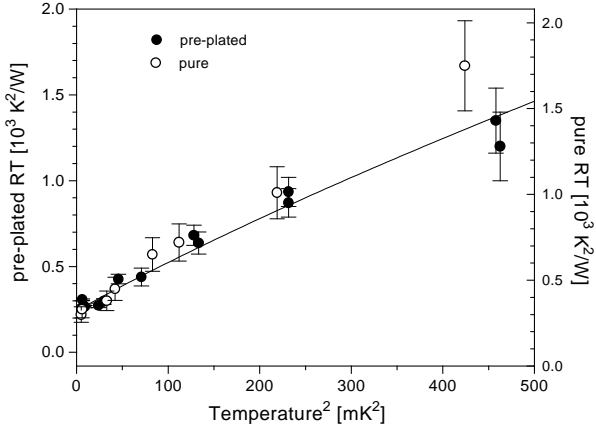}}
\caption{Graph of $RT$ vs. $T^2$ where $R$ is the measured thermal
resistance.  The closed circles (left scale) are the new results for the
$^4$He-plated plug.  The open circles (right, shifted scale) are the
earlier pure $^3$He results.  The fitted line is discussed in the text.}
\label{fig3}
\end{figure}

Fig.~\ref{fig3} is a plot of $RT$ versus $T^2$ for the present results
(closed circles, left scale) and for the earlier pure $^3$He results
(open circles, right scale). The $RT$ axes have the same scale factor
but have a relative shift; this was done to show that within the
experimental uncertainty there is a change in the intercept (the
boundary scattering term in the plug) but not to the slope (the
quasiparticle-quasiparticle term in the plug and in the bulk). The solid
line fit to the results is not in fact a straight line. Anderson et
al.\cite{43} and Greywall\cite{44} have measured the thermal
conductivity of bulk $^3$He and found corrections to the first order
result, $\kappa \propto T^{-1}$. We have used Greywall's result, $\kappa
= 2.91 \times 10^{-4} ~(T - 12.2T^2 + 74.4T^3)^{-1}$ W m$^{-1}$ K$^{-1}$
with $T$ in kelvin.

The intercepts are given by $RT = 240 \pm 20$ and $320 \pm 20$ K$^2$/W
for the four $^4$He monolayer and pure $^3$He measurements respectively,
a drop of 25\% on pre-plating. By using equation~(\ref{eq:9}) and an
effective area calculated with the model of Robertson et al.\cite{45},
the effective pore diameters were calculated to be $36 \pm 11$ and $27
\pm 8$~nm respectively.

The time constant for mechanical equilibrium of the $^3$He through the
porous plug was $7.0 \pm 0.1$ hours. This was a similar 30\% reduction
from the 10~hour time constant for the pure $^3$He experiment. The pore
diameter can be estimated from these time constants. Again we follow
Robertson et al.\ in assuming an intersecting cylinder model for the
pores. The volume flow rate along a cylinder of diameter $d$ and length
$L$, assuming diffuse scattering at the cylinder walls, is\cite{8}
\begin{equation}
{dV\over dt} = {\pi d^3\Delta P\over 4 n m^\ast v_{\rm F} L}
\end{equation}
where $n$ and $m^\ast$ are the number density and effective mass of the
$^3$He quasiparticles and $\Delta P$ is the pressure difference between
the ends. For a $(35 \pm 3)$\% packing fraction, the intersecting pore
model gives the following relation between the number of cylinders in
the flow direction and the cylinder diameter: $N\pi d^2/4 = (0.40 \pm
0.08)A$ where $A$ is the cross-section of the plug. A straightforward
mechanics calculation of the time constant then yields $d = 43 \pm 9$
and $30 \pm 7$~nm for the cylinder (pore) diameters for the four $^4$He
monolayer and pure $^3$He measurements respectively. 

Table~\ref{table:1} brings together the results from the two different
measurements on the two $^3$He samples. Also included are the diameters
deduced from the Knudsen flow of $^4$He gas through the plug at room
temperature and from the viscous flow of pure liquid $^3$He at 500~mK,
where $\ell_q \ll d$\cite{1}. The first conclusion to be drawn is that
the estimates based upon boundary scattering are significantly smaller
than the diameter deduced from the surface area of the porous plug.
Therefore the intersecting cylinder model is not appropriate for the
packed powder despite giving a good description of sintered metal
powder. The results suggest that a large fraction of the boundary
scattering is in the backward direction, so decreasing the mass and heat
flows. This may be reasonable given that all surfaces are convex within
a packed powder whereas the formation of necks by the sintering process
will give rise to both convex and concave surfaces in a sintered metal
powder. The second conclusion is that the calculated diameters are
larger for the $^4$He-coated and presumably specular surfaces. However,
without a good model for the structure, it is not possible to relate the
measured pore diameters to a specularity parameter.
\begin{table}[t]
\caption{The effective pore diameter of the porous plug determined by
different methods.  The liquid $^3$He fluid flow was performed below
20~mK in the boundary scattering regime.  The $^4$He gas measurement was
made at room temperature; the mean free path was $\sim$300~nm.  The heat
flow measurements were derived from the low temperature limit of the
thermal resistance.  Viscous flow measurements were performed at 500~mK
where $\ell_q \ll d$.  In all cases the Robertson et al.\ model was used
to describe the porous structure and accounted for most of the
uncertainty.}
\begin{tabular}{lccc} 
Experiment & \multicolumn{3}{c}{Pore Diameter (nm)} \\ \hline
Adsorption Isotherm &&90$\pm$8 & \\  \cline{2-4}
 & \multicolumn{2}{c}{Liquid $^3$He} & $^4$He Gas \\
		\cline{2-3}
 & Pure & $^4$He-Plated  & \\ \hline
Fluid Flow (Knudsen) & 30$\pm$7 & 43$\pm$9 & 40$\pm$8 \\ %\hline
Heat Flow & 27$\pm$8 & 36$\pm$11 &  \\
Viscous Flow (500 mK) & 80$\pm$20 & & 
\end{tabular}
\label{table:1}
\end{table}

\begin{figure}
\epsfxsize=3.2in
\centering{\epsfbox{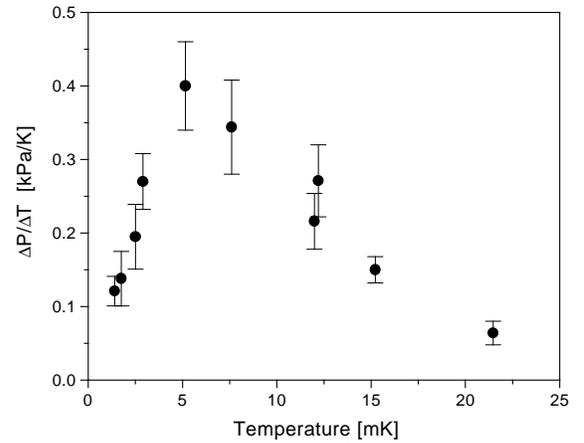}}
\caption{The measured liquid $^3$He thermomechanical coefficient 
for the $^4$He-plated
porous plug.}
\label{fig4}
\end{figure}

\begin{figure}
\epsfxsize=3.2in
\centering{\epsfbox{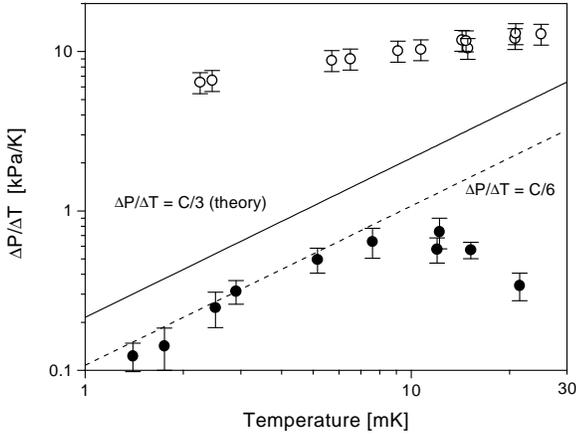}}
\caption{A log-log plot of the derived boundary-scattering liquid $^3$He
thermomechanical coefficient for the $^4$He-plated porous plug (closed
circles) as a function of temperature.  Also shown are the earlier pure
$^3$He results (open circles) and the theoretical result from Edwards et
al.\ (solid line).  The dashed curve is discussed in the text.}
\label{fig5}
\end{figure}

Fig.\ \ref{fig4} shows the measured thermomechanical coefficient as a
function of temperature. 
This coefficient is a blended combination of
two separate coefficients $(\Delta P/\Delta T)_{\rm bd}$ and $(\Delta
P/\Delta T)_{\rm qp}$ arising from scattering of the quasiparticles by
boundaries and other quasiparticles respectively. We follow the
Nordheim-Gorter approach to the thermoelectric power resulting from two
or more separate scattering processes and write,
\begin{equation}
\left( {\Delta P\over \Delta T} \right)_{\rm meas.} = {({\Delta P\over 
\Delta T})_{\rm bd}
R_{\rm bd} + ({\Delta P \over
\Delta T})_{\rm qp} R_{\rm qp}\over R_{\rm bd} + R_{\rm qp}}
\end{equation}
where $R_{\rm bd}$ is the thermal resistance due to boundary scattering
alone and $R_{\rm qp}$ is due to quasiparticle scattering alone. We know
that $(\Delta P/\Delta T)_{\rm qp} = 0$, within experimental
error\cite{1}. Furthermore the measured thermal resistance has been
separated as shown in Fig.~\ref{fig3}. The derived $(\Delta P/\Delta
T)_{\rm bd}$ is shown as solid circles in Fig.~\ref{fig5}, again as a
function of temperature. A log-log format was chosen so that the present
results could be compared with the earlier work on pure $^3$He (open
circles) and the theoretical calculation of Edwards et al.\cite{9}
(solid line).

Clearly, the $^4$He pre-plating had a striking effect on the
thermomechanical coefficient, reducing it by up to 30$\times$ at 2~mK.
This is to be compared to the $\sim$ 30\% effect on the thermal and flow
conductivities in the low temperature limit of boundary scattering.
Other features to be brought out are as follows: The coefficient is
linear in temperature, within experimental error, in the low temperature
limit, as predicted by the theoretical calculation and as expected from
the analogy with diffusion thermoelectricity. However, the results seem
to follow $(\Delta P/\Delta T)_{\rm bd} \sim s/6v$, the dashed line,
rather than the theoretical $s/3v$. The reason for this disagreement is
not known; the theoretical result is independent of whether the geometry
is that of cylinders or a model porous system and even of whether the
boundary scattering is elastic or inelastic\cite{9}. Another aspect of
the results that is puzzling is that the coefficient is linear in
temperature only in the low temperature limit where $d \ll \ell_q$;
either the boundary scattering thermomechanical coefficient decreases
with increase in temperature or the Nordheim-Gorter procedure is not
applicable in the region where $d \agt \ell_q$. Even 20~mK is well below
the Fermi temperature and so it is unlikely that the coefficient would
decrease. The Nordheim-Gorter procedure, which was designed to deal with
two or more impurity types, may not be appropriate to describe the
transition from the Knudsen limit to the hydrodynamic limit. This is
perhaps a theoretical challenge.

The most intriguing result is that in the original experiment with pure
$^3$He the thermomechanical coefficient was an order of magnitude larger
than with the pre-plated surface, and the temperature dependence was
$\sim T^{0.3}$ even in our low temperature limit. Similar behaviour has
been seen in thermoelectricity: Dilute alloys of noble metals with
transition metal solutes have shown giant thermopowers with non-linear
temperature dependence\cite{46,10}. These have since been attributed to
the Kondo effect and other magnetic effects in these dilute magnetic
alloys\cite{11}. The switching off of the `giant' thermomechanical
coefficient by replacing the magnetic $^3$He surface atoms with
non-magnetic $^4$He surface atoms is convincing evidence for the
magnetic origin of the enhancement. Further support comes from an
entropy calculation. The thermomechanical coefficient depends upon the
entropy, as does the thermoelectric power in a metal \cite{10} and the
phonon thermomechanical effect\cite{Andreev}. In the confined geometry
of the porous plug the surface $^3$He atoms contribute a large fraction
of the total entropy. If we postulate that there are 18 atoms/nm$^2$ in
the first two layers, an entropy of $k_B \ln 2$ per atom and use our
measured surface area (1.2~m$^2$) and pore volume $(2.5 \times
10^{-8}$~m$^3$), then $S \sim 200$~$\mu$J/K or $s/v \sim 8$~kJ/Km$^3$ or
$\Delta P/\Delta T \sim 8$~kPa/K. This calculation shows that the
degrees of freedom do exist in the magnetic solid layers to account for
a large thermomechanical coefficient. Measurements of the heat capacity
of the solid $^3$He magnetic layers on the surface of sintered silver
powder\cite{29} and vycor glass\cite{30} show a heat capacity of
$\sim$~15~$\mu$J/Km$^2$ from 1 to 20~mK and diminishing beyond 20~mK\@.
This heat capacity has been attributed to a wide range of Weiss
temperatures on the inhomogeneous surfaces. It would modify the entropy
calculation, dropping it by about 25\% at 2~mK and by a diminishing
fraction as temperature is increased. This could explain the weak
temperature dependence of the thermomechanical effect with pure $^3$He.

\section{Conclusions}

The addition of four monolayers of $^4$He to the surface of the porous
plug lowered the thermomechanical coefficient due to boundary scattering
by up to a factor 30 to a low temperature limiting value of $\Delta
P/\Delta T = (100 \pm 15)T$ kPa/K $\sim s/6v$, with $T$ in kelvin. This
is half the theoretical result, a disagreement for which we have no
explanation. Looking back to the original pure $^3$He result, it was
clearly anomalously high and reminiscent of the giant thermopowers in
dilute magnetic alloys. The dramatic effect of adding the non-magnetic
$^4$He monolayers is evidence that the enhancement was due to magnetic
scattering by the solid-like Curie-Weiss magnetic $^3$He monolayers at
the interface. The nature of the scattering is not clear. From the
discussion in the Introduction we know that scattering of the
quasiparticles by the magnetic $^3$He layers is diffuse. However,
replacing the $^3$He by $^4$He at the interface changes the scattering
to specular, signifying that, on the length scale $\sim 2\pi/k_{\rm F}$,
the scattering surfaces must be geometrically flat. It is also known
that both spin-flip and energy exchange, during boundary scattering of
the $^3$He quasiparticles, have a very low probability $(\sim 10^{-8})$
even for pure $^3$He; therefore the scattering is close to elastic in
the usual sense. The only possibility is to suppose that the pure $^3$He
interface appears magnetically rough to the $^3$He quasiparticles.

Within experimental error the measurements of fluid flow and heat flow
through the porous plug were consistent and gave effective pore
diameters of $\sim$~40~nm for the $^4$He pre-plated surface and
$\sim$~30~nm for the pure $^3$He case. The comparison is evidence for
increased specularity for the $^4$He-plated surface. However, since both
effective pore diameters are well below the 90 nm deduced from the
surface area of the plug, there is need for a better model to describe
transport in the pressed powder plug. As discussed above, probably a
large fraction of the $^3$He quasiparticle scattering was
back-scattering.

From the above conclusions, we can see a clear need for more
experiments. A more ideal geometry is needed to confine the $^3$He. The
Anopore material\cite{47} used for helium studies by Hallock's
group\cite{48} looks very promising; it has parallel non-intersecting
cylindrical pores with a high porosity. At the same time, the technique
needs to be revised to reduce the 7--10 hour time constant to achieve
equilibrium. Nevertheless, the giant thermomechanical effect has been
demonstrated and remains to be understood in terms of scattering by the
magnetically rough interface.

\section{Acknowledgments}

First we wish to thank Professor David Edwards and his group for their
interest in this work; it was the comparison between the original
measurements and the theoretical calculation by his group that led to
this study. We are grateful to Kim MacKinder, Steve Gillen and Jim
Thompson for technical support and to NSERC for financial support. We
have also benefited from discussions with Professors Robin Fletcher and
Eugene Zaremba. Finally, we thank Professor Mezhov-Deglin for bringing
reference\cite{Andreev} to our attention and for describing his attempts
to measure a phonon thermomechanical effect.

\end{document}